\documentclass[11pt]{article}

\newfont{\msbm}{msbm10}

\def\K{{\cal K}}

\def\A{{\cal A}}

\def\R{\hbox{\msbm R}}

\def\F{{\cal F}}

\newtheorem{prop}{Proposition}
\newtheorem{cor}{Corolary}
\newtheorem{theo}{Theorem}
\newtheorem{deff}{Definition}

\def\bc{\begin{cor}}
\def\ec{\end{cor}}

\def\bt{\begin{theo}}
\def\et{\end{theo}}

\def\bd{\begin{deff}}
\def\ed{\end{deff}}

\def\bp{\begin{prop}}
\def\ep{\end{prop}}

\def\ba{\begin{eqnarray}}
\def\ea{\end{eqnarray}}

\def\be{\begin{equation}}
\def\ee{\end{equation}}

\newfont{\msbms}{msbm6}  

\def\E{{\cal E}}

\def\ab{\mbox{$\bar \A$}}
\def\Si{\mbox{$\Sigma$}}

\usepackage{color}

\begin{document}
\title{Groups of  flux-like transformations \\
in loop quantum gravity}

\author{J. M. Velhinho}

\date{{ Departamento de F\'\i sica, Universidade da Beira 
Interior\\R. Marqu\^es D'\'Avila e Bolama,
6201-001 Covilh\~a, Portugal}\\{jvelhi@ubi.pt}}

\maketitle

\begin{abstract}
\noindent  
We present a group of transformations in the quantum configuration
space of loop quantum gravity
that contains the set of all transformations generated by the flux variables. 
\end{abstract}

\pagestyle{myheadings}


\section{Introduction}

Besides the configuration variables, associated to holonomies,
a key ingredient in  loop quantum gravity (LQG) \cite{AL,T} is the set of momentum  variables,
given by fluxes through appropriate surfaces.
The flux variables  correspond to  `infinitesimal translations' in the configuration
space, and the one-parameter groups of  transformations generated by each flux
variable are well known \cite{LOST,F2,BT}. However, the set of  transformations obtained by exponentiation of
the fluxes does not form a group.

Following our previous work \cite{Vf},  we present here
a group 
of transformations  that includes all of those generated 
by the flux variables. All transformations in this group 
leave the  Ashtekar-Lewandowski measure invariant,
and so they are all unitarily implemented
in the standard LQG  representation.
In what follows, we will focus on the trivial bundle case and on the analytic set-up 
(see  \cite{Vf,V,Vg} for details). 

\section{Preliminaries}

Let then \Si\ be an  analytic spatial manifold and $G$ a compact 
Lie group ($G=SU(2)$ in LQG).
Consider  oriented analytic curves
in \Si, or more precisely, 
the set $\E$ of equivalence classes known as {\em edges}, where two curves are identified if they differ only by a (orientation preserving)
reparametrization.

A generalized connection $\bar A$ is a map $\bar A:\E\to G$ such that
$\bar A(e^{-1})=\left(\bar A(e)\right)^{-1}$
and 
$\bar A(e_2 e_1)=\bar A(e_2)\bar A(e_1)$,
whenever the composition of edges $e_2 e_1$ is again an edge. (The composition
of edges is induced by the natural composition of curves.) 
The set of all such maps $\bar A$  is the space of generalized connections \ab, which
plays the role 
of a (kinematical) `quantum configuration' space.

 Let us introduce the notion of {\it germ}, for edges \cite{T}. 
Consider the  equivalence
relation in $\E$ defined by: two edges are equivalent if they start at the same point and  one of the edges is an analytic
extension of the other. An equivalence class of edges is called a germ.
So, a germ at a point $x$ is defined by the infinite set
of Taylor coefficients, from which
the whole family of equivalent edges can be reconstructed.
The set of germs at any given point $x\in\Si$ is independent
of $x$, and will be denoted by $\K$. The germ of $e$ is  denoted by $[e]$.

\section{Basic group}

Since $G$ is a group,  the set $\rm{Map}[\Si\times \K,G]$ of all maps $g:\Si\times \K\to G$ is a  group under the pointwise product: $(gg')(x,[e]):=g(x,[e])g'(x,[e])$.
We are interested in a subgroup thereof, defined as follows.
For a given edge $e$ and a point $x$ on $e$, let $e_x$ denote the subedge of $e$ that starts at $x$. Now, for $g:\Si\times \K\to G$ and an edge $e$, let $S(g,e)$ denote the set of points $x$ along the edge $e$ such that
 $g(x,[e_x])$ is different from the identity of the group $G$.
The subgroup $\F\subset \rm{Map}[\Si\times \K,G]$ is defined to be  
{\em the set of all elements $g$ such that 
$S(g,e)$ is a finite set for every given\/ edge $e$}.
 It is clear that $\F$ is a subgroup,
since if $S(gg',e)$ is an infinite set for some edge $e$, then either
$S(g,e)$ or  $S(g',e)$ (or both) must be infinite.

The following are examples of elements of the group $\F$.  

{\em Example 1:}  $g\in \rm{Map}[\Si\times \K,G]$ supported on a finite number of points, i.e. such that
$g$ is different from the identity of $G$ only in a finite number of points
of $\Si$.

{\em Example 2:} Elements $g\in \rm{Map}[\Si\times \K,G]$ supported
in an analytic (or semianalityc) surface $S$, and such that $g$ also equals the identity for every germ that defines curves on  the surface  $S$. Appart from that, $g$ may vary from point to point
on $S$ and depend on the germ at each point.
(These elements are well defined,  since
the number of points $x$ in the intersection
between an analytic edge $e$ and a (semi)analytic  surface $S$  such that 
$[e_x]$ does not define curves in  $S$ 
is  finite.)

A subgroup $\cal TF$ of the group $\F$ is obtained when we consider functions $g\in\F$ that depend
 only on the tangent direction of the germ, i.e. such that,
at each point, $g(x,[e])=g(x,[e'])$ whenever germs $[e]$ and $[e']$ have the same
tangent direction at the starting point $x$. 
The subgroup $\cal TF$
can be defined exactly like $\cal F$, with $\cal K$ being replaced by the set $S^2$ of  directions
in the tangent space  $T_x\Si$ at a point $x$.

Elements of $\cal TF$ directly related 
to the LQG flux variables  are defined as follows.

{\em Example 3.}
Let $S$ be an oriented
(semi)analytic surface and  g a $G$-valued function on $S$. An element $g\in\cal TF$ is obtained by declaring that $g$ is supported on $S$ and, on points $x\in S$:
{\it i)}  $g$  is the identity
for   every germ that defines curves on $S$; {\it ii)} $g(x,[e])={\rm g}(x)$ for germs (that define curves) pointing upwards (with respect to the orientation of $S$); {\it iii)} $g(x,[e])={\rm g}(x)^{-1}$ for germs  pointing downwards.


\section{Generalized flux transformations}


Our main result is that the group $\F$ can be seen
as a group of transformations in the space \ab.
In more precise terms, 
{\em there is a faithful representation, hereby denoted $\Theta$,
of   $\F$ as a group of transformations in \ab.} Moreover,
{\em the  transformations  generated by the LQG flux variables belong to the subgroup 
$\Theta (\cal TF)$} \cite{Vf}.

To define  the representation $\Theta$, it is sufficient to give, for each $g\in \F$,
the images  $\Theta_g(\bar A)$ of every $\bar A\in\ab$, which in turn are determined once $\Theta_g(\bar A)(e)$ is known for every edge. 
In order 
to
define $\Theta_g(\bar A)(e)$, for any given $g$, $\bar A$ and edge $e$, one can use the same procedure as for the construction
of graphs adapted to a given surface \cite{T}.
Essentially, given $g\in\F$ and an edge $e$, 
one works out a decomposition of the edge:
\be
\label{dec}
e=e_m^{\epsilon_m} e_{m-1}^{\epsilon_{m-1}}\cdots e_1^{\epsilon_1},
\ee
such that each subedge $e_k$ starts at a point of $S(g,e)$ and contains no other point of that
set. Here, the symbols $\epsilon_k$ can take the values $\pm 1$.

Once a decomposition of the type (\ref{dec}) is performed,  it suffices to give the images of the edges $e_k$
in the decomposition:
\be
\label{teta2}
\Theta_g(\bar A)(e_k)={\bar A}(e_k)\, g^{-1}(x_k,[e_k]),\ \ \ k=1,\ldots,m,
\ee
where   $x_k\in S(g,e)$ is the starting point of the edge $e_k$.
The images $\Theta_g(\bar A)(e)$ are then determined by
$\label{teta1}
\Theta_g(\bar A)(e)=[\Theta_g(\bar A)(e_m)]^{\epsilon_m} [\Theta_g(\bar A)(e_{m-1})]^{\epsilon_{m-1}}\cdots [\Theta_g(\bar A)(e_1)]^{\epsilon_1}$.

It is straightforward to check that $\Theta$ is  a representation,
and so the group $\F$ can indeed be interpreted as a group of transformations in the space $\ab$. To see that $\F$ contains the transformations associated with the LQG fluxes, let us consider elements $g\in{\cal TF}$ of the type given in
Example 3, with the $G$-valued functions ${\rm g}(x)$ being generated by Lie($G$)-valued
functions $f(x)$, i.e. ${\rm g}(x)=\exp(tf(x))$, $t\in\R$. The   transformations $\Theta_g$ corresponding to these elements are precisely the transformations in \ab\ generated by the LQG flux variables (see Eqs. (17) in \cite{LOST}). 

An important result  is that, just like the transformations generated by the fluxes, 
{\em all transformations in the  group $\Theta(\F)$ are continuous with respect to the natural
topology of \ab, and leave the Ashtekar-Lewandowski measure invariant} \cite{Vf}.
(The proof of this fact is essentially the same as in the case of the 
flux variables \cite{LOST,F2,OL}).
It follows
that the standard LQG representation of the holonomy-flux algebra carries an unitary representation of the group  $\Theta(\F)$,
which extends the set of unitary operators obtained by 
quantization of the exponentiated fluxes.





\end{document}